\documentstyle[twocolumn,prb,aps,epsf]{revtex} 
 
\begin{document} 
\input epsf 
 
\wideabs{ 
\title{Correlated atomic motion and spin-ordering in bcc $^{3}$He}   
\author{N. Gov $^1$ and E. Polturak $^2$}   
\address{$^1$Department of Physics, 
University of Illinois at Urbana-Champaign, \\ 
1110 Green St., Urbana 61801, IL, USA \\ 
$^2$Physics Department, 
Technion-Israel Institute of Technology,\\ 
Haifa 32000, Israel}   
\maketitle   
\tightenlines  
\widetext  
\advance\leftskip by 57pt  
\advance\rightskip by 57pt

\begin{abstract}   
We propose a new way to treat the ordering of nuclear magnetism of solid $^{3}$He.  
We argue that the magnetic interaction arises indirectly as a consequence of correlated   
zero-point motion of the ions. This motion   
lowers the energy of the ground state,   
and results in a coherent state of oscillating electric dipoles. Distortion of the electronic 
wavefunctions leads to hyperfine magnetic interactions with the nuclear spin. 
 Our model describes   
both the modification of the phonon spectra, localized modes ("vacancies") and the  
 nuclear magnetic ordering of bcc $^{3}$He using a single parameter, the  
 dipolar interaction energy $E_{0}$. The model yields correctly both the u2d2 symmetry   
of the ordered phase and the volume dependence of the magnetic interaction.  
We calculate the magnetic excitations in the u2d2 phase, which compare well with the measured specific-heat, free energy and entropy. 
We further give a description of the nature of the High-Field phase and the phase-transition lines as a function of magnetic field and volume. 
\end{abstract}   
   
\vskip 0.2cm  
} 
 
\narrowtext  
\tightenlines  
\vspace{.2cm} 
 
\section{Introduction} 
 
The spin-ordered phase of bcc $^{3}$He presents a difficult challenge to 
accurate theoretical description \cite{fisher,cross}. The main problem is to 
explain why the transition temperature \cite{halperin} of 10$^{-3}$K is two 
orders of magnitude larger than the nuclear dipolar interaction $\sim $10$%
^{-8}$K. The prevalent description is in terms of atomic exchange cycles 
involving several atoms (Multiple Spin Exchange model). Successively higher order cycles 
 produce competing ferro and antiferromagnetic interactions \cite{roger}
For qualitative description 
within this Multiple Spin Exchange model (MSE), one 
needs up to six or more atoms exchanging places, which becomes 
difficult to calculate and seems a priori less probable \cite{fisher,grey1}. 
In fact, it is not sure that this description converges \cite{cross,ceperley,cowan,kumar} 
for larger exchange cycles, i.e. may not be the spin-ordering mechanism.
This conceptual problem, despite the relative successes of the MSE picture \cite{osherof88}, therefore raises the interest in 
another approach \cite{qfs}.

We propose that magnetic 
ordering of $^{3}$He is a result of correlations in the zero point atomic motion, which 
however does not require exchange of atoms \cite{niremil}. These 
correlations lower the energy of the ground state \cite{nirbcc} and result 
in a coherent state of oscillating electric dipoles, which also modify the phonon 
spectrum. 
We show here that the zero-point motion of the atoms produces an oscillating 
magnetic polarization 
of the electronic cloud which interacts with the nuclear spin in each atom. 
The nuclear spins interact with each other indirectly,  through the electric dipolar 
interaction between the atoms. In effect, 
 this interaction is hyper-fine like, and has the right order of magnitude to be 
related to the spin-ordering transition. In addition, we show that the volume dependence 
of the on-site magnetic interaction is in agreement with experimental data, and that 
the space symmetry of the proposed interaction leads naturally to the 
distinct u2d2 antiferromagnetic phase. Our scenario 
is somewhat similar to the mechanism of hyperfine enhanced nuclear magnetism 
that occurs in certain insulators \cite{suzuki}.

The plan of the paper is as follows: 
In section II we establish the applicability of our model of zero-point electric dipoles 
in bcc $^{3}$He. 
In section III we describe the hyperfine magnetic interactions and their volume dependence. 
In section IV we show how the indirect dipolar interactions induced by the 
on-site hyperfine interactions lead to the u2d2 symmetry of the ordered phase. 
In section V we calculate some properties of the u2d2 phase and compare them to the 
experimental data. 
In section VI we give a description of the nature of the high-field phase (HFP) 
and calculate its properties. 
 
\section{Coherent dipoles in the bcc ground-state and transverse phonons} 
 
At temperatures which are high compared to the magnetic interactions (T$\gg $%
1mK), the atomic motion in bcc $^{3}$He can be treated in the same way as in $%
^{4}$He \cite{niremil}. 
 
We begin by observing \cite{niremil,glyde} that in contrast to the situation prevailing in 
gaseous He, the interatomic potential is highly anisotropic 
in the bcc crystal. The potential well is both very wide and anharmonic (double-well) along 
the major axes (100,010,001). 
This means that the atomic wavefunctions will be particularly extended in these directions. 
In order to reduce the repulsive overlap energy, the many-body wavefunction has to 
include "dynamic-correlations" \cite{nosanow}. 
The current treatment using variational wavefunction \cite{glyde} incorporates these 
correlations using a Jastrow-type wavefunction that has a two-body or higher order terms 
to represent the short-range hard-core repulsion. 
Using this variational wavefunction, the phonon spectrum is calculated through a 
Self-Consistent Harmonic (SCH) calculation \cite{glyde}. This gives a satisfactory 
description of most phonon branches, except for the transverse T$_{1}$(110) phonon, which  
experimentally is found to be much softer than calculated. 
Introducing cubic terms into the calculation to represent the anharmonicity, softens the 
calculated phonon spectra overall, 
which ruins the good agreement with the phonon branches other than T$_{1}$(110). 
This SCH+C method also breaks the delf-consistency of the Harmonic procedure and predicts an anomalous upward slope of the lowest transverse T$_{110}$ mode, which is not observed.
 
We would like to describe the anharmonicity in a different way; we treat the local motion 
of the atom in its potential well as a "local mode". The harmonic description of the crystal 
potential misses the low-lying vibration of the atoms due to the 
shallow (double-well) potential along the major axes, which is the 
mode responsible for the strong dynamic correlations in the many-body 
wavefunction. We therefore treat the low-energy local excitation along the 
major axes, and the harmonic modes as two independent degrees of freedom. 
The collective motion of the atoms 
is assumed to be well described by the Self Consistent Harmonic calculation. 
In this  procedure, we take the SCH spectrum from previous 
works \cite{glyde}, as the approximate harmonic description. 
The "local modes" can be shown to couple only to the T$_{1}$(110) transverse 
phonon\cite{niremil}. The coupling is described in terms of an electric 
dipolar matrix element,  
where this zero-point dipole moment arises from a correlated 
atomic motion along the major axes. 
 
The origin of a zero point electric dipole moment is the highly
directional zero point oscillation of the nucleus, which breaks the rotational symmetry of 
the nuclear position relative to the electronic cloud. 
Going beyond the Born-Oppenheimer approximation, the electronic wave
function aquires a p-component in addition to the s-orbital.
The first order correction
to the electronic energy due to 
relative nuclear-electronic fluctuations is \cite{vibronic}: 
$\Delta E \simeq (m/M_n)(a/D)E_{sp} \simeq 5$K, where $m$ is the electron mass, $M_n$ the 
nuclear mass, $a\simeq0.25$\AA \ is the Bohr radius, $D\sim0.8$\AA \ is the spread of the excited
nuclear wavefunction, and $E_{sp}$ the excitation energy of the He atom from the $s$ to 
the $p$ shell. 
Since the crystalline potential is anisotropic, we can take this zero point motion to be 
highly directional (i.e. a completely off-diagonal matrix element). 
We assume that $\Delta E$ is the energy 
associated with the mixing of
the $s$ and $p$ electronic levels: $\left| \psi \right\rangle \simeq \left| s\right\rangle 
+\lambda \left| p\right\rangle $, i.e. $\lambda \sim \sqrt{\Delta E/E_{sp}}\sim 10^{-2}$. 
This local polarization results in 
zero-point dipolar interactions between the atoms \cite{niremil,nirbcc}. Taking these 
interactions into account, 
the ground state of the crystal can be lowered by having the nuclear motions 
correlated with respect to the relative phase of this zero-point oscillatory motion. The dipolar 
interaction energy $E_{0}$, of the order of $\Delta E$, is the energy we associate 
with the correlated zero-point oscillations of the atoms along the major axes. 
 
In our scenario, the ground-state of the 
crystal in which the zero-point motion of the atoms is correlated may be 
described as a global state of quantum resonance 
between the two degenerate configurations shown in Fig.1, each of 
which minimizes the instantaneous dipolar interaction energy \cite{niremil} 
for dipoles along the major axes (i.e. ${\bf \hat{\mu} }={\bf \hat{x} },{\bf \hat{y} },{\bf \hat{z} }$)
\begin{equation} 
E_{dip}=-\sum_{i\neq 0} {\bf {\mu_{0}\cdot \mu_{i}}}\left[ \frac{3\cos%
^{2}\left( {\bf \hat{\mu} }\cdot \left( {\bf r}_{0}-{\bf r}_{i}\right) \right) -1}{%
\left| {\bf r}_{0}-{\bf r}_{i}\right| ^{3}}\right]  \label{edip} 
\end{equation} 
where ${\bf r}_{0}$ is the position of an atom and the summation ${\bf r}_{i} 
$ runs over the entire lattice. The atomic electric dipole moment $\left|  
{\bf \mu }\right| $ induced by the coherent zero-point motion is proportional to $%
\lambda $ through  
\begin{equation} 
\left| {\bf \mu }\right| =e\left\langle \psi \left| x\right| \psi 
\right\rangle \simeq 2e\lambda \left\langle s\left| x\right| p\right\rangle  
\label{mu0} 
\end{equation} 
 
Due to the lower symmetry of 
the dipolar array (Fig.1) compared to the crystal symmetry, a coupling between the 
harmonic phonons and the dipolar modulation exists only along the (110) 
direction \cite{niremil}. The excitations of these dipoles are periodic modulations of 
the relative phases of the dipoles, 
with respect to the ground state array (Fig.1). We found that the dipolar 
interactions vanish at the edge of the Brillouin zone, 
so this transverse phonon has the energy $E_{0}$ at this point. 
The hybridization of the dipolar excitations with the phonons results in a factor 2 softening of the transverse T$_{1}$(110) phonon 
compared to the harmonic calculation. The experimental data give a softening by a 
factor of $\sim$1.7, in close agreement. 
In addition, in our model there appears a localized (i.e. dispersionless) 
excitation branch of energy $2E_{0}$, which is involved in mass diffusion and 
contributes to the specific-heat \cite{niremil}. This excitation is a quantum analogue of a 
point defect ("vacancy") and has similar features \cite{niremil}. 
The specific heat of the localized 
mode is consistent with the experimental contribution 
attributed to point defects to the total heat capacity, and in fact 
resolves previous inconsistencies related to the properties of the point defects 
in this solid (vacancies)\cite{goodkind}. 
Let us emphasize that the additional modes are a consequence of introducing an 
additional degree of freedom to the solid \cite{nirbcc}, which is the relative phase of 
the internal mixing of the $\left| s\right\rangle$ and $\left| p\right\rangle$ states. 
 
In the case of bcc $^{4}$He the bare local mode energy $E_{0}$ was measured 
directly by NMR of a $^{3}$He impurity, and the phonon spectrum was measured 
using neutron scattering. Our calculation is consistent with both sets of 
data \cite{niremil}. There are no neutron scattering measurements of the 
phonon spectrum of bcc $^{3}$He, 
but sound velocity data 
indicates that the slope of the T$_{1}$(110) phonon is about half of the SCH 
calculation \cite{greywall,kohler}. We predict that this ratio should 
indeed be 1/2, and we therefore take half the energy of the calculated \cite{kohler} SCH 
T$_{1}$(110) phonon at the edge of the Brillouin zone to be the bare 
dipole-flip energy $E_{0}(\simeq 5$K at V=21.5cm$^{3}$/mole) in solid $^3$He. According to our 
model, the energy of the localized mode involved in thermally activated 
self-diffusion should be $2E_{0}\simeq 10$K. This value is in excellent agreement 
with the activation energy measured by x-ray diffraction, ultrasonics and 
NMR experiments \cite{heald} at V=21.5cm$^{3}$/mole. A similar activation 
energy is also obtained from the excess specific heat \cite{grey}, and 
pressure measurements \cite{izumi}. 
 
We therefore establish the likely occurance of coherent zero-point dipoles in the 
ground-state of bcc $^{3}$He. 
The volume dependence of the dipolar energy (i.e. the 'vacancy' activation energy 
$2E_{0}$) is shown in Fig.2a from various experimental sources \cite {heald,greywall}. 
Since the excitation energy of the atom inside the crystal potential increases with decreasing 
volume, we indeed expect the energy $2E_{0}$ to behave as in Fig.2a. 
The mixing parameter $\lambda $ depends on pressure through $\sqrt{VE_{0}}$, 
where $V$ is the molar volume. This behavior arises from the definition of $%
E_{dip}$ (\ref{edip}),(\ref{mu0}) and the factor $\left| {\bf r}_{0}-{\bf r}%
_{i}\right| ^{3}$ in this equation is proportional to the molar volume $V$. 
We plot the result in Fig.2b, where we see that $\lambda $ increases slowly 
with decreasing molar volume $V$. 
Future x-ray studies \cite{simmons} in bcc $^{3}$He may be able to check our prediction 
for the relation between the activation energy ($2E_{0}$, Fig.2a) and the T$_{1}$(110) 
phonon as a function of the molar volume. 
 
\section{Magnetic interactions} 
 
We now describe the magnetic interaction arising within our model. The lowest $%
\left| p\right\rangle $ level of the He atom has the electrons in a spin $S=1 
$ state due to strong exchange interaction \cite{cohen}, of the order of 
0.25eV. 
The zero-point electric dipole interactions, arising from the relaxation of the 
Born-Oppenheimer approximation, are also accompanied by electronic spin interactions due 
to magnetic-dipole 
transitions. The size of the electric dipole matrix is given by the electric dipolar 
interaction energy 
$E_{0}=2|E_{dip}|$ (\ref{edip}). 
The size of the magnetic-dipole matrix is related to the electric dipole matrix 
by \cite{cohen} 
\begin{equation} 
E_{magtrans}\simeq E_{0} \cdot (\hbar/M_ac_{s})/\left\langle x \right\rangle \label{emag}  
\end{equation} 
where $M_a$ is the atomic mass, $c_{s}$ is the velocity of sound of the T$_{110}$ phonon. 
We find that $\hbar/mc_{s} \sim \left\langle x \right\rangle\simeq0.8$\AA \ so that both 
transition matrices are of the same order.                                               
We therefore have oscillating zero-point electric and magnetic dipoles, with an electric 
dipole moment of size $\sim 2e\lambda \left\langle x \right\rangle$ 
and a magnetic dipole moment of size $\sim 2\lambda n_e$, where $n_e$ is the electron's 
magnetic moment. 
 
In addition, the atomic $p$-state, $S=1$ level is split into 3 sublevels with quantum 
numbers: $J=L+S=0,1,2$, 
due to spin-orbit coupling \cite{cohen}, which is of the order of $\sim 1.5$K. 
In our particular $sp$-hybridized states, which are not pure states of the quantum number $J$, 
there is no spin-orbit coupling, as the orbital current is zero. Rather, 
these states are split as a consequence of the magnetic dipolar interaction between the 
two electrons. The magnetic interaction 
will tend to align the total electronic spin with the symmetry axis of the 
excited $\left| p\right\rangle$ state, i.e. along the normal axes (Fig.3). 
This follows from a simple calculation of the magnetic dipolar interaction 
of the two electrons, in which the $\left| s\right\rangle $ electron is centered on 
average in the nucleus, while the $\left| p\right\rangle $ electron is spread over the 
two lobes (Fig.3). We thus estimate the aligning dipolar interaction energy at 
$E_{es}\sim 1.4$K, of the order of the spin-orbit coupling. 
 
In the ground state, the zero-point $\left| p\right\rangle $ electrons 
will therefore have an oscillating magnetic moment  
${\bf N}_{e}\sim 2\lambda n_{e}$, where $n_{e}$ is the magnetic 
moment of an electron. Because part of the electronic magnetic moment is now 
associated with the $%
\left| p\right\rangle $ state, there appears a net uncanceled moment of 
equal size in the $\left| s\right\rangle $ component of the electronic 
wavefunction. In $^{4}$He the nuclear spin is zero, and the zero-point magnetic polarization 
of the electrons has no effect. In contrast, in $^{3}$He the nuclear 
magnetic spin $I=1/2$ will interact with the oscillating electronic magnetic 
moment, mainly due to the contact term of the $\left| s\right\rangle $ 
electron at the nucleus. The magnetic interaction is of the hyper-fine type  
\cite{cohen}, and the energy associated with it, $E_{mag}$, is given by  
\begin{equation} 
E_{mag}=\left\langle -\frac{8\pi }{3}{\bf N}_{e}\cdot {\bf N}_{n}\delta 
(r)\right\rangle   \label{edipcont} 
\end{equation} 
where ${\bf N}_{n}$ is the effective nuclear magnetic moment, reduced empirically \cite{grey1} 
by a factor of $0.83$ due to zero-point fluctuations of the nuclear spins \cite{xxzmodel}. 
Using experimental hyper-fine data \cite{hyperx}, 
we find the maximum value of $E_{mag}/k_{B}\simeq 4.5$mK (for V=24 cm$^{3}$/mole). 
This energy is much larger than the direct nuclear dipole-dipole interaction ($\sim 0.1\mu$K), 
and is of the right magnitude to explain the high transition temperature of nuclear 
ordering in bcc $^{3}$He. A diagram of the energy scales involved in the magnetic interactions 
is shown in Fig.4, where we show the cascade of 
the different interactions that ultimately align the electronic and nuclear spins, 
with the zero-point excitations of the $p$-state.

Before proceeding to describe the nuclear spin ordering phenomenon, and the effect 
of the magnetic interactions on the zero-point electric dipoles, 
we present a quantitative calculation of the volume dependence of the 
magnetic interaction at T=0. 
It is known empirically that these interactions have a strong volume 
dependence \cite{fisher}, usually described as: $E_{mag}\propto V^{18}$. 
In our model, the 
magnetic energy (\ref{edipcont}) will change with pressure due to changes of 
the electronic magnetic polarization ${\bf N}_{e}$. First, the pressure 
changes the admixture of the $\left| p\right\rangle $ orbital in the ground 
state wavefunction $\left| \psi \right\rangle $, through changes of the 
mixing parameter $\lambda $. Second, as the solid is compressed, the 3 
sub-levels ($J=0,1,2$) broaden into partially overlapping bands. This 
overlap leads to further reduction of the net alignment of the electronic 
magnetic moment ${\bf N}_{e}$ with the orbital polarization (Fig.3). 
 
The reduction in ${\bf N}_{e}$ due to the broadening of the 
sub-levels with pressure can be approximated as linearly proportional to the overlap integral 
of the three spin-orbit sub-levels  
\begin{equation} 
\left\langle {\bf N}_{e}(V)\right\rangle \simeq \left( 1-F(V)\right) \lambda 
(V){\bf N}_{e}  \label{magf} 
\end{equation} 
 
where the overlap integral $F(V)=\int \phi _{1}^{\ast }(E)\phi _{-1}(E)dE$, 
and $\phi _{1}^{\ast }(E)$,$\phi _{-1}(E)$ are taken as normalized 
Gaussians centered at the energies of the spin-orbit levels, 
which are $\sim 1.5$K apart \cite{cohen}. The effective 
energy width $\gamma $ of these Gaussians ($\phi _{1}^{\ast }(E)$ and $\phi _{-1}(E)$) is 
estimated using a simple band calculation, as the change of the Coulomb energy 
of the $\left| p\right\rangle $ electrons due to overlapping $\left| 
p\right\rangle $ wavefunctions on neighboring He nuclei \cite{ashcroft}  
\begin{equation} 
\gamma (V)=\lambda ^{2}\sum_{i}\int \frac{e^{2}}{\left| {\bf r}_{0}-{\bf r}%
_{i}\right| }\psi _{0}^{\ast }(r)\psi _{i}(r)d^{3}r  \label{width} 
\end{equation} 
 
where $\psi _{0}^{\ast }(r),\psi _{i}(r)$ are the $p$-state wavefunctions of 
the electrons around the central atom and its $i$ neighbors. In this 
calculation we also take into account the large spatial spread of the atomic 
position inside the wide potential well \cite{niremil}, of approximately $%
\pm 0.8$\AA . We find that $\gamma $ ranges from $\sim $0K for V=24 cm$^{3}$%
/mole to 5.5K for V=19 cm$^{3}$/mole. The overlap factor $1-F(V)$ is therefore very 
sensitive to volume, changing from 1 at V$\sim24.3$ cm$^{3}$/mole to $\sim 0.01$ at 
V=19 cm$^{3}$/mole. One can see that as the volume decreases the broadening 
of the bands increases, thereby decreasing the net magnetic polarization 
of the electronic cloud. 
 
The strength of the magnetic interaction should be proportional to the 
measured Curie-Weiss temperature $\theta $. In Fig.5 we compare the normalized magnetic 
interaction energy $E_{mag}$ (\ref{edipcont}) with the normalized 
measured \cite{fisher} $\theta $. We find that the volume dependence of $%
E_{mag}$ agrees very well with that of $\theta $. 
Since in our model there is only one magnetic interaction, there is a single energy scale that 
determines the volume dependence of the transition temperature and the $e_2$ coefficient 
of the specific-heat \cite{fuku91}. 
All these quantities have the same volume dependence, a simpler situation than that 
involving a competition among many exchange processes \cite{fisher,osherof88}. 
 
\section{Composite Quantum Resonance of the electronic and magnetic subsystems}
 
We now consider the effects of the magnetic interactions on the coherent-state of 
the oscillating electronic dipoles. We begin with the high-temperature paramagnetic 
phase (PP), where the nuclear spins are randomly oriented. The existence of 
the hyper-fine splitting $E_{mag}$ (\ref{edipcont}) means that the quantum resonance 
state of the atoms, oscillating between the two configurations of Fig.1 is broken. 
The resonance frequency of each site will be slightly different from each other, 
randomly shifted by $0-E_{mag}/\hbar$, along each of the 3 major axes. The result will be 
that the zero-point electric dipoles will slowly drift out of resonance, breaking the 
long-range order of Fig.1. 
 
It is possible to restore the degeneracy of the overall ground state, 
and hence the quantum resonance condition of the zero-point dipolar oscillations. This can 
be done if the nuclear 
spins become ordered as well, in a spatial configuration which has the same overall 
magnetic interaction energy with each of the two states of Fig. 1. 
We first note that the total on-site magnetic interaction is isotropic with respect to 
the direction of the nuclear spin. This is due to the simultaneous interaction of the 
nuclear spin with the zero-point electronic spin which has equal components oscillating 
along all 3 major axes. 
 
The electric 
dipolar interactions in the ground-state of Fig.1 are such that each simple 
cubic sublattice of the bcc has no net interaction with the other 
sublattice. We therefore look for the possible static arrangements of the nuclear spins that 
fulfill the resonance requirement on each sublattice independently. Such arrangements are 
shown in Fig.6. In these arrangements there is an equal 
number of atoms with electronic and nuclear spins aligned (and anti-aligned) 
in both degenerate configurations of the electric dipoles (Fig.1). This 
arrangement also preserves the overall time-reversal symmetry of the system at 
zero field, i.e. no net magnetic polarization. Since these configurations exist on both 
sublattices, we end up 
with the u2d2 arrangements which is the symmetry of the ordered nuclear phase  
\cite{osheroff,bossy}. 
The only constraint so far due to the quantum resonance is the u2d2-type symmetry 
(i.e. also u3d3 etc.). 
We would like now to describe in more detail the nature of the coherent 
state combining the zero-point electronic dipoles with the ordered nuclear spins. 
 
The random on-site magnetic interaction can be thought of as 
a fliping of the dipoles with respect to the ground-state configuration of Fig.1. 
The average density, or probability, of a dipole-flip is given by the ratio of the 
magnetic energy $E_{mag}$ to the electric dipole-flip excitation energy $E_{0}$. 
The ground-state occupation of these flipped dipoles due to the magnetic interactions, 
means that there is a change in the zero-point occupation of the $p$ electrons $\lambda$. 
The $p$-state mixing which is associated with these magnetically-induced flipped 
dipoles, is given by: $\lambda_{m}\sim \lambda \sqrt{E_{mag}/E_{0}}\simeq 10^{-2}\lambda$. 
This small polarization corresponds to an electric dipole moment: 
${\bf \mu }_{m} \simeq 10^{-2}{\bf \mu}$, where ${\bf \mu}$ is the full zero-point 
electric dipole moment (\ref{mu0}). 
Since the magnetically-induced electric dipole-moments ${\bf \mu }_{m}$ are interacting 
with each other, 
we are looking for configurations (of the type shown in Fig.6) that minimize this 
dipolar interaction energy (\ref{edip}) along the major axes (i.e. ${\bf \hat{\mu} }={\bf \hat{x} },{\bf \hat{y} },{\bf \hat{z} }$)
\begin{equation} 
E_{magdipole}=-\sum_{i\neq 0} {\bf {\mu_{m,0}\cdot \mu_{m,i}}}\left[ \frac{3\cos%
^{2}\left( {\bf \hat{\mu}_m }\cdot \left( {\bf r}_{0}-{\bf r}_{i}\right) \right) -1}{%
\left| {\bf r}_{0}-{\bf r}_{i}\right| ^{3}}\right]
\label{edipolem} 
\end{equation} 
 
The configuration of Fig.6a has the lowest dipolar energy given by (\ref{edipolem}), and will 
therefore be chosen as the ground-state of the ordered phase. 
In this configuration, the nuclear spins are aligned along a vector ${\bf d}$ which is 
orthogonal to the u2d2 modulation axis ${\bf l}$ (here along the (100) axis). One 
can see that the total dipolar interaction energy will be invariant under any rotation 
of the vector ${\bf d}$ around ${\bf l}$. 
We therefore naturally obtain the state where the vector ${\bf l}$ is along one of 
the major axes, as is found experimentally from the domain structure \cite{osheroff}. 
We point-out that the u2d2 symmetry of the ordered phase results from symmetry and energy 
considerations which are independent of any quantitative parameters. This is in contrast 
to the MSE theory where the symmetry is determined by competition between many 
parameters \cite{fisher,roger}. 
 
In our picture, the interaction between the nuclear spins is indirect, 
being mediated by the electronic subsystem. This picture is somewhat 
similar to the hyperfine enhanced nuclear magnetism observed in certain 
insulators \cite{suzuki}. There, the nuclear polarization is effectively enhanced 
by an electronic spin that is partially aligned by a strong hyperfine 
interaction of a charged rare-earth ion. In our case, it is a quantum 
resonance of the electronic system that results in a coherent phase of 
oscillating electric and magnetic moments. 
 
Experimentally, there is a marked difference between the hcp and bcc phases, in the sense 
that there is spin-ordering in the bcc phase, 
as opposed to the lack of such a transition in the hcp phase \cite{okamoto}. This 
result is a natural consequence of our model, since we expect the triangular symmetry 
to frustrate any long-range dipolar ordering \cite{niremil} of the type shown in Fig.1. 
Indeed there is no evidence for long-range zero-point electric dipole correlations in the hcp 
phase \cite{niremil}, in the form of large discrepancies between the SCH calculations 
of the phonon spectrum and the measured phonons. 
We therefore do not expect a priori any long-range ordering of the nuclear spins in the 
hcp phase \cite{hcpdipole}. In contrast, 
within the conventional MSE model, the absence of ordering of the spin system well below its 
Curie-Weiss temperature is not easily explained.

\section{Magnetic excitations and energy of the U2D2 phase} 
 
Due to the large ratio between $E_{0}$ and $E_{mag}$, the 
slow zero-point oscillations of the magnetically-excited electric dipoles ${\bf \mu }_{m}$ 
can be decoupled from the much faster zero-point oscillation of the electric dipoles 
with natural frequency $E_{0}/\hbar $. The new coherent state can therefore 
be described as a tensor product of the coherent states of the two decoupled 
components of the oscillating electric dipoles. A fast component with 
natural frequency $E_{0}/\hbar $ described by the Bose creation/annihilation 
operators $a{{_{k}}^{\dagger }}$,$a_{k}$ representing approximately the full electric 
dipole moment (\ref{mu0}), and a slow component described by the Bose creation/annihilation 
operators $b{{_{k}}^{\dagger }}$,$b_{k}$ representing the electric dipole 
moment ${\bf \mu }_{m}$ induced by the on-site hyperfine magnetic interaction. 
This decoupling of the magnetic and 
phononic quasi-particles is quite natural \cite{fisher}, due to the difference 
of 3 orders of magnitude in the energy. 
 
If we treat the magnetically-induced interactions as decoupled from the fast oscillating 
component, the total effective Hamiltonian for the new coherent state is 
\begin{equation} 
H=H_{phon}+H_{mag}  \label{htot} 
\end{equation} 
where $H_{phon}$ is given in Eq.5 of Ref.[\cite{niremil}], and similarly 
\begin{eqnarray} 
H_{mag}&=&\sum_{k}\left( 2E_{magdipole}+X_{mag}({\bf k})\right) b_{k}^{\dagger 
}b_{k}  \nonumber \\ 
&&\ \ +\sum_{k}X_{mag}({\bf k})\left( b_{k}^{\dagger }b_{-k}^{\dagger 
}+b_{k}b_{-k}\right)  \label{hmag} 
\end{eqnarray} 
where $b_{k}^{\dagger },b_{k}$ are Bose creation/annihilation operators that 
represent a local flip of ${\bf \mu }_{m}$ 
out of the ground-state (u2d2) arrangement, which behaves as a local-mode 
with bare energy (\ref{edipolem}) 
$E_{magdip}\equiv 2E_{magdipole}\simeq E_{mag}$, i.e. $X_{mag}(0)=-E_{magdip}/2$. 
The dipolar interaction $X_{mag}({\bf k})$ is given by (\ref{edipolem})\cite{niremil} 
\begin{eqnarray} 
X_{mag}\left( {\bf k}\right) &=&-\sum_{i\neq 0} {\bf {\mu_{m,0}\cdot \mu_{m,i}}}
\left[ \frac{3\cos ^{2}\left( {\bf \hat{\mu }}_{m}\cdot \left( {\bf r}_{0}-{\bf r%
}_{i}\right) \right) -1}{\left| {\bf r}_{0}-{\bf r}_{i}\right| ^{3}}\right] \nonumber \\ 
&&\  \ \times \exp \left[ 2\pi i{\bf k}\cdot \left( {\bf r}_{0}-{\bf r}_{i}\right) \right] 
\label{xmag} 
\end{eqnarray} 
with ${\bf k}$ being the wavevector characterizing the 
modulation of the array of the dipoles ${\bf \mu }_{m}$ and nuclear spins, 
arranged in u2d2 configuration Fig.6a. Since the relative 
phase between the magnetically-induced electric dipole moments ${\bf \mu }_{m}$ is determined by 
the relative alignment of the nuclear spins, 
there is a one-to-one correspondence between the modulation of the nuclear spins relative to  
the u2d2 arrangement, and modulation of the relative phase of the electric dipoles 
${\bf \mu }_{m}$. 
The local-mode of energy $E_{magdip}$ therefore corresponds also 
to a nuclear spin-flip out of the u2d2 ground-state arrangement. 
 
The Hamiltonian ${H_{mag}}$ (\ref{hmag}) which describes the effective 
interaction between the magnetically induced dipoles ${\bf \mu }_{m}$ can be diagonalized 
using the Bogoliubov 
transformation \cite{anderson} ${\beta _{k}}=u(k)b_{k}+v(k)b{^{\dagger }}_{-k}$. The two 
functions $u(k)$ and $v(k)$ are given by: 
$u{^{2}}(k)={\frac{1}{2}}\left( \frac{E_{magdip}{+X}_{mag}{(k)}}{{E}_{sw}{(k)}}%
+1\right) ,v{^{2}}(k)={\frac{1}{2}}\left( \frac{E_{magdip}{+X}_{mag}{(k)}}{{E}%
_{sw}{(k)}}-1\right)$, where the modulations of the magnetic Hamiltonian ${H_{mag}}$ describe 
the energy spectrum of the spin-waves. 
\begin{equation} 
E_{sw}(k)=\sqrt{E_{magdip}\left( E_{magdip}+2X_{mag}(k)\right) }  \label{ek} 
\end{equation} 
The result of solving the effective 
Hamiltonian (\ref{hmag}) is 
a coherent ground-state \cite{huang,nirbcc} 
\begin{equation} 
\left| \Psi _{0}\right\rangle= 
\left( \prod_{k}\exp \left( \frac{v_{k}}{u_{k}}b_{k}^{\dagger }b_{-k}^{\dagger }\right) \right) 
\left| vac\right\rangle   \label{psi0} 
\end{equation} 
 
A modulation  of the relative phases of the magnetically-induced 
electric dipoles ${\bf \mu }_{m}$ with a wavevector ${\bf k}$, represents a spin-wave 
since the nuclear spins have 
the same relative phase. We see that the electric dipolar interaction introduces an 
effective restoring force between the nuclear spins, that results in the 
above dispersion relation. 
 
In Fig.7 we plot the dispersion relations of the spin waves obtained from Eq. (\ref{ek}), along 
various directions of the crystal. We find 
that the velocities of these waves at long wavelengths are in the range $%
6.0-10.0$cm/sec (for V$\simeq 24$cm$^3$/mole). 
These values are similar to those extracted from melting pressure measurements \cite{osherofyu}, where a value of $8.4\pm 0.4$cm/sec was found. 
There is a soft excitation in the (100) direction ($c_{sw}\sim 3$%
cm/sec), and also an unstable excitation with an imaginary dispersion relation in the 
(010) direction, along which we therefore do not expect a propagating 
spin-wave. Along this direction the modulation in the nuclear spins breaks the quantum-resonance 
condition, which states that the total magnetic interaction energy is the same for both 
configurations of the dipoles of Fig.1. This modulation is therefore prohibited within the 
quantum resonance restriction. 
We note that this calculation was for the nuclear spins fully aligned with the $z$-axis. 
 
Mean-field calculations of the MSE model \cite{sun} have also resulted in anisotropic 
spin-waves, though there is at present no experimental data on the directional 
anisotropy of the spin-waves to compare with theory. 
 
In addition to Bosonic spin-waves there can also be local dipole-flips of the 
magnetically-induced electric dipoles ${\bf \mu }_{m}$. These $\pi$-shifted, i.e antisymmetric, dipole excitations are treated as Fermions, 
just as 
described previously for the full electric dipole moment \cite{nirbcc}. The energy spectrum 
of these excitations is given by \cite{nirbcc} 
\begin{equation} 
E_{fm}(k)=\sqrt{E_{magdip}^{2}+E_{sw}(k)^{2}}  \label{specdirm} 
\end{equation} 
The dispersion relation of these optic-like excitations is also plotted in Fig.7. 
 
We can now calculate the contribution of all these excitations to the specific-heat. 
In the calculation, we included the 
individual contributions of each mode (in a spherical phase-space volume), and then simply 
averaged. Each type of excitation was treated with using the appropriate (Bose or 
Fermi) statistics, and the details are described in the Appendix (\ref{freeu2}). 
The agreement with the (limited) experimental data for 
the volume V=24.13 cm$^3$/mole\cite{grey1} is fairly good (Fig.8). 
At this molar volume, we have $E_{magdip}\simeq 2.6$mK. We note that 
the contribution of the optic-like modes is important at higher temperatures, and that the 
agreement between the data and the simple $T^3$ law is somewhat fortuitous. Similar 
conclusions were reached using the MSE model \cite{sun}. 
In particular, our calculation has the right slope as the experimental data. 
 
For the specific volume $V\simeq24$cm$^3$/mole, experimental data on the ground-state 
energy \cite{fuku87,fuku91} gives $E_{mag}/2\sim 1.1\pm 0.1$mK, while our 
calculation (\ref{edipcont},\ref{magf}) gives $E_{mag}/2\sim1.3$mK. Plotting the free 
energies for the two phases (\ref{freeu2}),(\ref{freepar}) in Fig.9, we find that they 
give a transition temperature: $T_{N}\sim1.25$mK, close to the measured result \cite{fuku87} 
of $\sim1.05$mK. 
In this calculation, we took the empirical value of $E_{mag}/2=1.1$mK, for the constant 
aligning energy appearing in (\ref{freeu2}),(\ref{freepar}). In addition, 
the experimental data in Fig.9 was shifted up by $0.1$mK, which is within the experimental 
uncertainty. 
We find that the flat free energy below the transition temperature fits the free energy of 
the ordered phase (\ref{freeu2}). It should be pointed out that this calculation, which is based 
on independent spin-waves, is valid only at low temperatures, compared to $E_{mag}$. 
We see that the energy difference between different spin arrangements which fit the symmetry 
constraint (such as u3d3 etc. \cite{fisher}), is only through the relatively small 
term $\Delta E_{dip}$ in the free-energy (\ref{de0}). 
 
In Fig.10 we show  the entropies of the two phases as a function of temperature, and 
indicate the jump at the approximate transition temperature. We obtain a latent 
heat of $\Delta S T_{c} \sim 100\mu$K per atom, while experimentally it is found to 
be \cite{grey1} $\sim 250\mu$K per atom. 
This degree of agreement is quite reasonable considering the crudeness of the 
spin-wave approximation 
at temperatures where the entropy is already $\sim 0.35 \ln (2)$, and the independent 
spin-wave picture is questionable. 
 
Finally we describe the magnetic response of the u2d2 phase. The zero-temperature 
susceptibility of an antiferromagnet with the field orthogonal to the magnetic 
moments \cite{kittel}, is given by: 
$\chi(0)=\mu_{n}^2/E_{mag}$, where $E_{mag}/2$ is the aligning energy, and we use 
the effective nuclear moment, reduced by a factor of $0.83$, due to the quantum 
fluctuations \cite{grey1}. The calculated susceptibility is smaller by $\sim 15\%$ than the 
measured value \cite{osherof87}, and is therefore in reasonable agreement. 
 
The zero-field antiferromagnetic resonance frequency \cite{cross}, measured by 
NMR \cite{osheroff}, is given by: $\hbar^2 \Omega_{0}^2=4E_{d}E_{mag}$, where $E_{d}$ is 
the direct nuclear magnetic dipolar interaction. 
To calculate the temperature dependence of this quantity, we calculate (Appendix) the 
reduction in the ordered magnetic moment due to thermal occupation of excited 
states \cite{fisher}. 
Since $E_{d}$ depends on the ordered nuclear magnetic moment $N_{n}$ squared, we have 
that $\Omega_{0}\propto N_{n}$. 
In Fig.11 we plot the normalized temperature dependence of $\Omega_0$ compared to the 
thermally reduced magnetic moment: $(1-\Delta M)$ (Eq.(\ref{spinred})). We find that the 
reduction in the frequency is well described by our calculation, up to $\sim 0.7$mK, above 
which the experimental data have a steeper decline. 
We further note that NMR experiments measure the resonance of the stationary nuclear 
spin, and do not reflect the fast-oscillating ($\omega=E_{0}/\hbar$) zero-point electronic 
magnetic moments ${\bf N}_{e}$. 
 
Our final comment about the u2d2 phase is concerning its domain structure. It was 
found \cite{osherof90,sasaki} that a crystal grown in a narrow channel will have one 
large magnetic domain which will be oriented such that the magnetic planes are parallel 
with the channel axis and surface. 
This observation can be explained by our model in the following way. On the channel surface 
the coherent motion of the atoms is inhibited in the direction normal to the wall. The only 
remaining coherent motion is parallel to the wall surface, and so will therefore be the 
magnetic ordering. 
At the nucleation site, when the crystal grows into the free liquid, it will have the three 
domains along the three orthogonal major axes. The relative sizes of the domains will be 
determined by random roughness at the nucleation site \cite{sasaki}. 
 
\section{High-Field Phase} 
 
Above a critical external magnetic field $H_{c1}$ there is a first-order transition to a 
high-field phase (HFP). 
The NMR data for this phase \cite{hfp} indicates a phase with cubic symmetry and large 
magnetization: $N_{z}>0.6$ (in units of the saturation magnetization \cite{osherof87}). 
The nuclear spin arrangement infered for this phase is shown in Fig.12. 
Since there is now global nuclear magnetization along the field direction ($z$-axis in Fig.13), there 
is no time reversal symmetry and no quantum resonancea, for the oscillating electric dipole moments ${\bf \mu }_{m}$ 
along this axis. 
Along the orthogonal $xy$-directions there is still quantum resonance for the electronic 
dipoles ${\bf \mu }_{m}$'s, and hyperfine interactions of energy $E_{mag}$. 
In a semi-classical picture, the nuclear spins are vectors with the magnetization in 
the $z$-direction given by their $z$-component, precessing around the $z$-axis with the 
Larmor frequency. 
In the paramagnetic-phase (PP) the $xy$-components of the spins are random, while in the HFP there is long-range 
order for these components (Fig.13), due to the aligning effect of the hyperfine interactions 
in the $xy$-plane. 
The order-parameter is therefore the global phase of the coherent $xy$-components ${\bf \mu }_{m}^{xy}$, 
i.e. a complex order-parameter, and we expect a second-order phase transition of the 
type found for example in superfluid $^4$He, which is also in the 3D-XY universality class \cite{domb}. 
Further, the specific-heat data \cite{grey1} of the HFP has a distinct cusp at the transition temperature, typical of a $\lambda$-type transition.

Let us examine the field dependence of the hyperfine magnetic interactions in 
the $xy$-plane, by first looking at the electronic polarization. 
With increasing external magnetic field the electronic spins ${\bf N}_{e}$ in the $xy$-plane 
will become oriented with the external field along the $z$-direction, reducing the 
hyperfine interactions with the nuclear spins in the $xy$-directions (\ref{edipcont}). 
The electronic spins are oriented with the oscillating $p$-states of the atom in the $xy$ 
plane, with effective energy $\left\langle E_{es} \right\rangle \sim 1.4(1-F)$K, where the factor $1-F$ takes into account the 
overlap of the different electronic spin orientations (\ref{magf}, Fig.3), reducing the effective energy gap. The 
single-atom magnetic Hamiltonian for the electronic spins in the $xy$-plane, in an external field $H$ is therefore 
\begin{equation} 
H_{elec}\simeq -\frac{1}{2} E_{es} {N_{e}^{x}}^2 -H N_{e}^{z} 
\label{elecspin} 
\end{equation} 
with $N_{e}^{x},N_{e}^{z}$ the classical components of ${\bf N}_{e}$.
The first term represents the aligning dipolar interaction between the $s$ and $p$ electrons of the atom. 
Solving (\ref{elecspin}) by mean-field, we get a linear magnetization: $N_{e}^{z}=H/E_{es}$, and full magnetization of the electronic spins ${\bf N}_{e}$ at a field: $H_{c2}\sim 26$T (using $(1-F)\sim0.6$ for V$\sim24$cm$^3$, so that $E_{es}/2\sim0.4$K), in rough agreement with the estimated \cite{osherof87,osherof88} experimental value of $\sim22$T. 
At fields larger than $H_{c2}$ there are no $xy$ hyperfine interactions so the nuclear spins are free to fully align with the external field, and there is a zero-temperature transition to the PP. 
 
The volume dependence of $H_{c2}$ is experimentally \cite{fukuyama} found to have a volume dependence which is steeper than the critical Neel temperature $T_{N}$ of the u2d2 phase.
This is in agreement with our formulation (\ref{elecspin}), as $H_{c2}$ is given by the steep dependence of the factor $(1-F)$ (Fig.5). 
Since $T_{N}$ is roughly proportional to $E_{mag}$ (Fig.9), 
we get: $\frac{H_{c2}(V_1)}{H_{c2}(V_2)} \frac{\lambda(V_1)}{\lambda(V_2)}\simeq \frac{T_{N}(V_1)}{T_{N}(V_2)}$ (using Eqs.\ref{edipcont},\ref{magf} and $\lambda(V)$ from Fig.2a). 
Comparing the data \cite{osherof87,fuku91,fukuyama} at V$_2$=24 cm$^3$/mole and V$_1$=22.7 cm$^3$/mole we find: $\frac{H_{c2}(V_1)}{H_{c2}(V_2)}\sim0.25$, $\frac{\lambda(V_1)}{\lambda(V_2)}\sim1.2$ and $\frac{T_{N}(V_1)}{T_{N}(V_2)}\sim0.3$, in excellent agreement with the above relation. 
 
If the electronic spins ${\bf \mu }_{m}$ in the $xy$-plane would 
have an arrangement where the dipolar interaction energy is negative (such as the u2d2 phase (\ref{edipolem})), 
they would have zero-point oscillations with the resulting dipolar interaction energy $E_{magdip}$ (\ref{hmag},\ref{xmag}).
These oscillations would get them out of phase with the nuclear spins, which are precessing with the Larmor frequency. We therfore can only have a cubic arrangement with zero ground-state dipolar interactions.
In Fig.13 we plot the dipolar interaction energy $X_{hfp}$ (\ref{xmag}) of the $xy$-components ${\bf \mu }_{m}^{xy}$ as a function of modulation wavevector $k$ along the (001)-axis, with respect to the ground-state array of Fig.13.
We find that the dipolar interaction energy is indeed zero in the ground-state, with a quadratic dispersion similar to excitations in a ferromagnet. 
 
We shall now describe the nuclear spins in the HFP.
In the HFP the nuclear spins will have a stiffness against small perturbation from the ordered configuration of Fig.13 (see Fig.13). 
Perturbations of the nuclear spins in the $xy$-plane will cost an energy $X_{hfp}$ (Fig.13), while the $z$-component perturbation costs the Zeeman energy $E_{zeeman}=N_{n} H$. 
A simple description of this state is therefore by the following model magnetic Hamiltonian of independent nuclei
\begin{equation} 
H_{hfp}\simeq -H\sigma_z N_{n}^{z} -E_{x}\sigma_x N_{n}^{x} 
\label{ehfp} 
\end{equation} 
where $H$ is the external field in units of energy, $E_{x}=E_{mag}/2$ is the effective aligning energy of the $xy$-components 
compared with the random arrangement of the nuclear spins (see Eq.\ref{freepar}), 
$N_{n}^{z},N_{n}^{x}$ are the components of the nuclear spin.
The size of the hyperfine $xy$-interactions of the nuclear spins with the electronic spins will be reduced with increasing magnetic field: $E_{x}\propto {\bf N}_{e}^{x}\simeq \sqrt{1-H/H_{c2}}$ (\ref{edipcont},\ref{elecspin}). 
 
The energy of the nuclear spin is therefore given by $E_{hfp}=-\sqrt{H^{2} + E_{x}^2}$. 
We plot in Fig.14 the energy $E_{hfp}$ compared with the energy of the u2d2 phase at zero temperature (\ref{freeu2}): $E_{u2d2}=-(E_{mag}/2)-\Delta E_{dip}-\chi(0) H^2$ (using at V=24cm$^3$ $E_{mag}/2\sim1.1$mk, see Sec. V). 
We see that the HFP has lower energy than the u2d2 phase at fields $H>0.7$T, which is quite close to the measured value \cite{osherof87,osherof88} of $H_{c1}\sim0.45$T. 
In Fig.14 we also plot the magnetization of the u2d2 phase: $N_{u2d2}=\chi(0) H$, compared with the measured magnetization \cite{osherof87,osherof88}, and as calculated for the HFP (\ref{ehfp}): $N_{hfp}=-2H\left( H+E_{hfp} \right)/E_{x}^2$. 
We find that the calculated magnetization is $\sim 0.55$ at the calculated critical field $H_{c1}$, which is close to the measured value \cite{osherof87,osherof88} of $\sim 0.6$. 
 
We further checked the dependence of this critical field on the magnetic energy $E_{mag}$, and we find that it has a weaker than linear dependence, in accordance with the experimental data \cite{fuku91}. 
Taking $E_{mag}\propto T_N$ and using the measured \cite{fuku91} values at $V_{1}=24.21$cm$^3$ and $V_{2}=22.45$cm$^3$, we calculate a ratio of critical fields of $H_{c1}(V_{1})/H_{c1}(V_{2})\sim3.4$, compared to the measured ratio of $\sim3$ (while the transition temperatures vary by a factor of $\sim4.2$). 
 
We note that our description of the HFP is at best qualitative, since at higher magnetic fields the calculated magnetization $N_{hfp}$ is much larger than the measured magnetization (Fig.15). 
Using the measured magnetization we shall now use (\ref{ehfp}) to describe the phase diagram of the HFP.
The overall stiffness of the spins should be proportional to the transition temperature of HFP to PP, since the stiffness represents the energy cost to randomize the spins and create a PP. 
From (\ref{ehfp}) and Fig.13 this energy should be given by 
the geometric-mean of the energies appearing in (\ref{ehfp}): $E_{mean}=\sqrt{(H N_{z})(E_{x} N_{x})}$, where $N_{z}$ is the measured magnetization (Fig.15) and $N_{x}=\sqrt{1-N_{z}^2}$. 
We find that a constant factor of $\sim1.3$ relates this last expression ($E_{mean}$) with $T_{c}$, and is in excellent agreement with the measured $T_c$ (Fig.16).
 
\section{Conclusion} 
 
To conclude, our model enables us to describe both the phonon spectra and 
the nuclear magnetic ordering of bcc $^{3}$He using a single parameter, 
which is the thermal activation energy $E_{0}$. 
The model offers a consistent picture of the nature of the magnetic phases of bcc 
$^{3}$He, and their phase diagram. 
We give a quantitative description of the u2d2 phase, which compares well with the measured data. 
Despite the more qualitative nature of our description of the HFP, we obtain a good description 
of the phase diagram parameters: $H_{c1},H_{c2},T_{c}$ and their field and volume dependence. 
 
We are currently exploring the consequences of this model regarding the description of 
the magnetic properties of solid $^{3}$He in confined geometries (nano-clusters, aerogel) 
and in lower dimensions such as adsorbed two-dimensional films. 
 
{\bf Acknowledgements} 
N.G. thanks Tony Leggett and Gordon Baym for many useful discussions and suggestions. 
This work was supported by the Israel Science Foundation, the Technion 
VPR fund for the Promotion of Research, 
the Fulbright Foreign Scholarship grant, NSF grant no. DMR-99-86199  and 
NSF grant no. PHY-98-00978. 
 
\section{Appendix} 
Here, we present details of the calculation of the free-energy and magnetization 
of the u2d2 phase. 
 
The ground-state energy of the nuclear spins has two contributions. The first is the 
reduction of energy compared to the random case, where the electronic dipoles ${\bf \mu }_{m}$ 
of everage nergy $E_{mag}/2$ are excited. 
This energy is equivalent to an effective magnetic field that tends to align the nuclear 
spins, and its effects already appear significant in the PP. 
The free energy of the PP, can be therefore written as \cite{patria} 
\begin{equation} 
F_{para}\simeq -k_{B}T \ln \left[ 2 \cosh (E_{mag}/2k_{B}T) \right] \label{freepar} 
\end{equation} 
 
In the ordered, coherent u2d2 phase, 
there is an additional reduction in the ground-state energy of the coherent 
magnetically-induced electric dipoles ${\bf \mu }_{m}$ due to their negative dipolar interaction \cite{anderson,nirbcc} 
\begin{eqnarray} 
\Delta E_{dip}&=&\sum_{k}\frac{E_{sw}(k)-(E_{magdip}+X_{mag}(k))}{2} \nonumber \\ 
&&\simeq -50\mu K/atom  \label{de0} 
\end{eqnarray} 
where we again calculated the contribution from each Bosonic mode and then averaged. 
In the u2d2 phase there are therefore coherent zero-point electronic 
dipoles ${\bf \mu }_{m}$, which reduce the ground-state energy (\ref{psi0}), oscillating 
with a resonance energy $E_{magdip}$ (\ref{hmag},\ref{xmag}). 
The free energy of the ordered phase can therefore be approximated as 
\begin{eqnarray} 
F_{u2d2}&\simeq& \Delta E_{dip}-(E_{mag}/2) \nonumber \\ 
&&+k_{B}T \left[ \sum_{k} \ln \left( 1-\exp(-E_{sw}(k)/k_{B}T) \right) \right. \nonumber \\ 
&&\ \ \left. -\sum_{k} \ln \left( 1+\exp(-E_{fm}(k)/k_{B}T) \right) \right] \label{freeu2} 
\end{eqnarray} 
 
The specific-heat $C_{v}=-T\partial^2F/\partial T^2$ and entropy $S=\partial F/\partial T$ are then calculated from these free energies. 
 
We now calculate the reduction in the ordered magnetic moment due to thermal occupation of 
excited states \cite{fisher}. 
For the Bosonic and Fermionic excitations (\ref{ek},\ref{specdirm}) we have a reduction in 
magnetization which is proportional to the occupation of excited states 
\begin{eqnarray} 
\Delta M&=&\sum_{k}\left[ u_{k}^2 n_{B}+v_{k}^2(1-n_{B}) \right] \nonumber \\ 
&&\sum_{k}\left[ w_{k}^2 n_{F}+y_{k}^2 (1+n_{F}) \right] \label{spinred} 
\end{eqnarray} 
where $n_{B},n_{F}$ are the Bose/Fermi occupation factors respectively, and $(u_{k},v_{k}), 
(w_{k},y_{k})$ are the Bose and Fermi Boguliubov factors \cite{nirbcc}. 
The factors $(u_{k},v_{k})$ where given before (section V), while the Fermi factors are 
given by \cite{nirbcc}
\begin{equation} 
w_{k}^{2}=\frac{1}{2}\left( 1+\frac{E_{magdip}}{E_{fm}(k)}\right) ,y_{k}^{2}=\frac{1}{2} 
\left( 1-\frac{E_{magdip}}{E_{fm}(k)}\right) 
\label{uvdir} 
\end{equation}

\begin{figure}[tbp] 
\centerline{\ \epsfysize 5cm \epsfbox{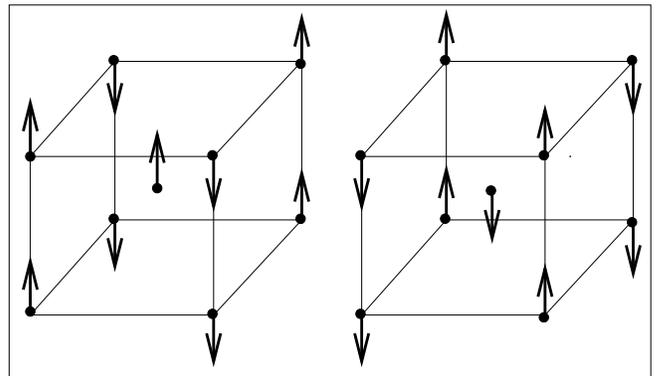}} 
\caption{The two degenerate 'antiferroelectric' dipole arrangement in the 
ground-state of the bcc phase. The arrows show the instantaneous direction 
of the dipoles.} 
\end{figure} 
 
\begin{figure}[tbp] 
\centerline{\ \epsfysize 10cm \epsfbox{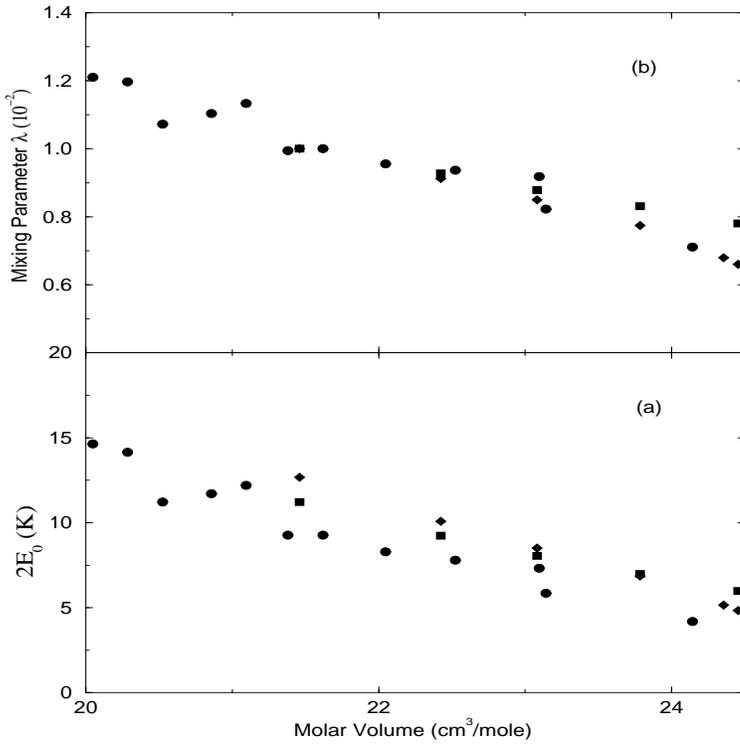}} 
\caption{(a) The dipolar activation energy $2E_{0}$ from experimental data: Circles \cite{heald}, squares \cite{grey} and diamonds \cite{izumi}. 
(b) The mixing parameter $\lambda$ (Eq.2) from the experimental data of (a).} 
\end{figure} 
 
\begin{figure}[tbp] 
\centerline{\ \epsfysize 5cm \epsfbox{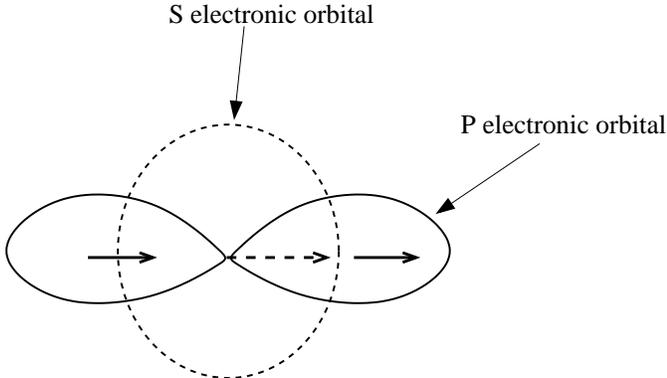}} 
\caption{The spin-orbit electronic magnetic-dipole interaction $E_{es}$ which aligns the $s$-shell spin in the center (dashed arrow) with the $p$-shell spin in the lobes (solid arrows).} 
\end{figure} 
 
\begin{figure}[tbp] 
\centerline{\ \epsfysize 5cm \epsfbox{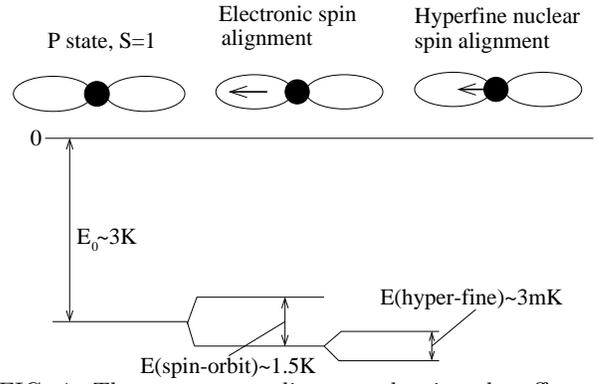}} 
\caption{The energy tree-diagram, showing the effects on the electronic and nuclear subsystems due to the coherent zero-point dipoles. The first effect is the excitation of $\lambda$ $p$-state with $S=1$, with energy $E_{0}$. The second effect is the alignment of the electronic spins with the electric dipole ($E_{es}$). The last effect is the hyperfine interaction that aligns the nuclear spin with the electronic spin with energy $E_{mag}$ (Eq.4).} 
\end{figure} 
 
\begin{figure}[tbp] 
\centerline{\ \epsfysize 7cm \epsfbox{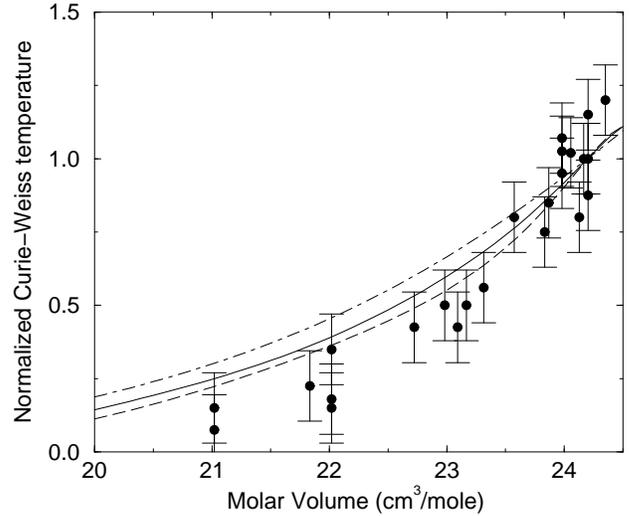}} 
\caption{The normalized magnetic splitting $2E_{mag}$ (Eq.4) as a function of volume, 
calculated using different experimental data sets for the activation energy $%
E_0$: dashed line \cite{heald}, dash-dot \cite{grey} and solid line \cite{izumi}. The solid circles are the normalized experimental Curie-Weiss temperature $\protect\theta$ \cite{fisher}.} 
\end{figure} 
 
\begin{figure}[tbp] 
\centerline{\ \epsfysize 5cm \epsfbox{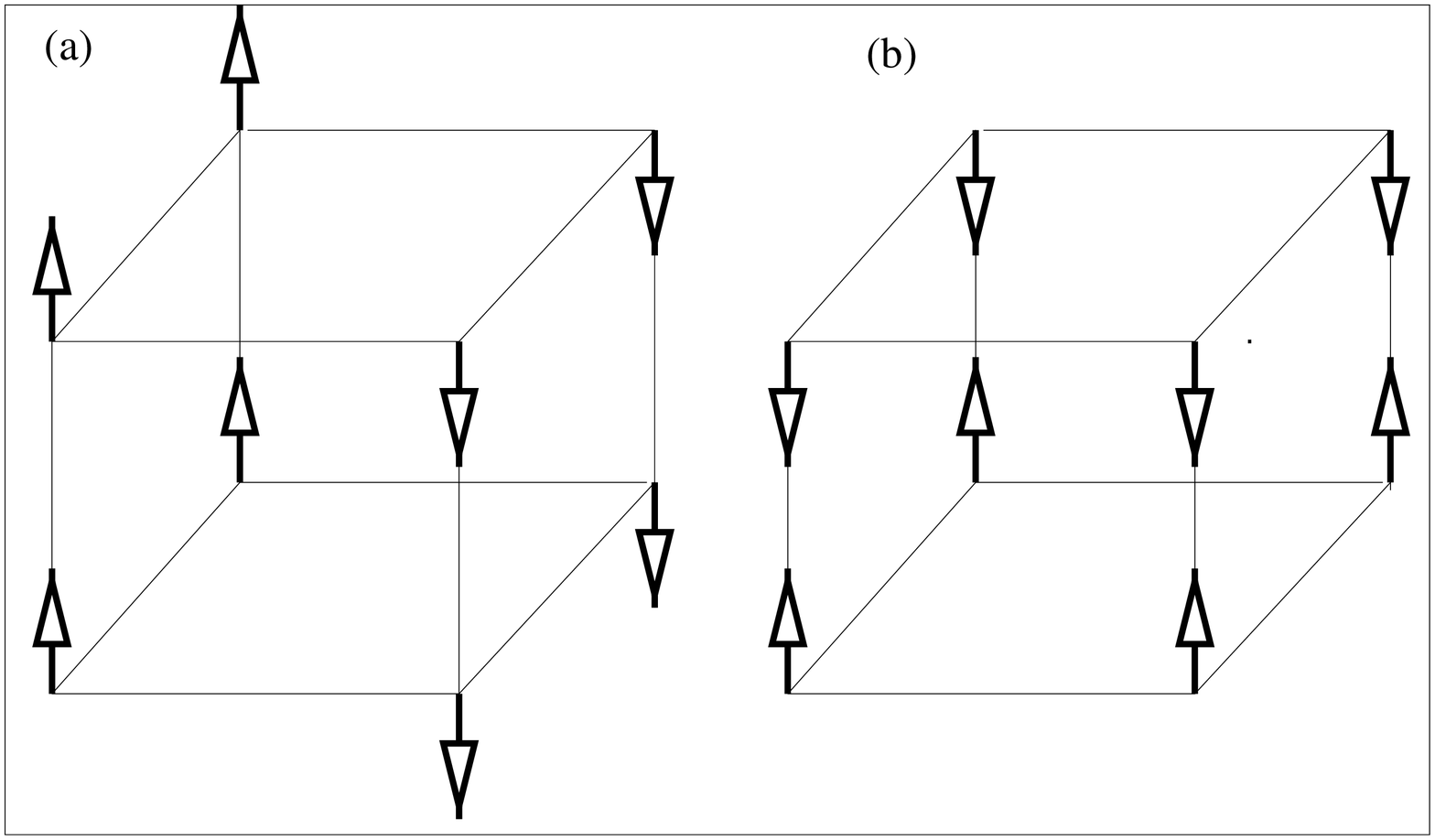}} 
\caption{Example of two simple u2d2 static nuclear spin (arrows) arrangements 
that maintain the quantum resonance of both the electric and magnetic 
dipoles (Fig.1). Only the spins on one simple-cubic sublattice of the bcc 
are shown, with the other sublattice having identical arrangement.} 
\end{figure} 
 
\begin{figure}[tbp] 
\centerline{\ \epsfysize 7.5cm \epsfbox{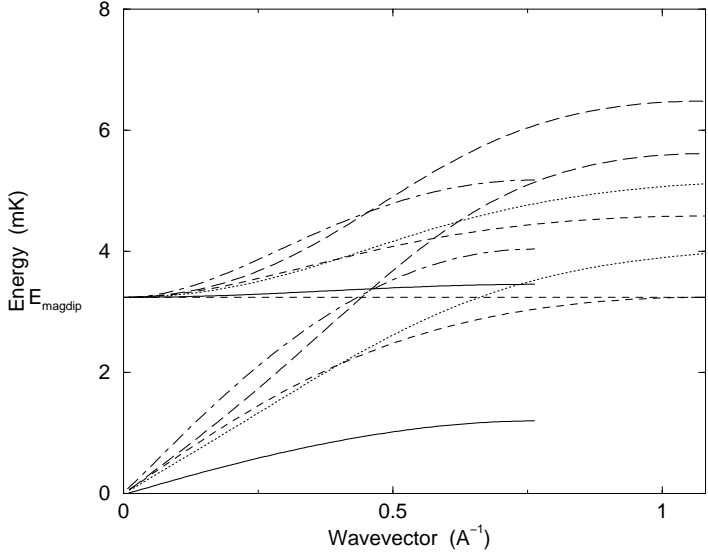}} 
\caption{The spin-wave spectrum $E_{sw}(k)$ (Eq.11) and the optic-like branches of the fermionic excitations $E_{fm}(k)$ (Eq.13). The nuclear spins are arranged as shown in Fig.6a, and the excitations are long the directions: (100)-solid, (101)-long dashed, (001) dash-dot, (011)-dashed, (111)-dotted.} 
\end{figure} 
 
\begin{figure}[tbp] 
\centerline{\ \epsfysize 7.5cm \epsfbox{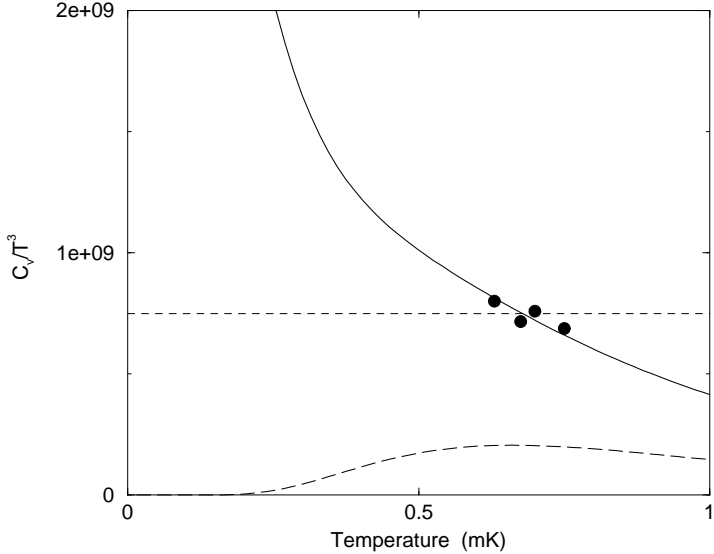}} 
\caption{The calculated total specific-heat (from Eq.18) (solid line), compared to experimental results \cite{grey1} at $V=24.13$cm$^3$/mole (solid circles). We also plot the contribution of the fermionic modes (heavy dashed-line), and the linear spin-wave approximation (thin dashed line). In this calculation we used $E_{mag}=2.6$mK.} 
\end{figure}                                                                      
 
\begin{figure}[tbp] 
\centerline{\ \epsfysize 7.5cm \epsfbox{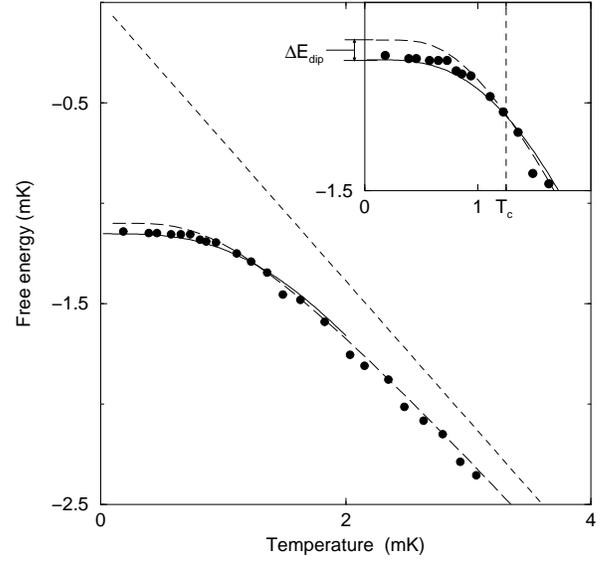}} 
\caption{The calculated free energies of the coherent u2d2 magnetic dipoles (solid line, Eq.18), of the PP (dashed line, Eq.16), of free spins (thin dashed line) and the experimental data \cite{fuku87} (solid circles). The inset shows the transition region.} 
\end{figure}  
 
\begin{figure}[tbp] 
\centerline{\ \epsfysize 6cm \epsfbox{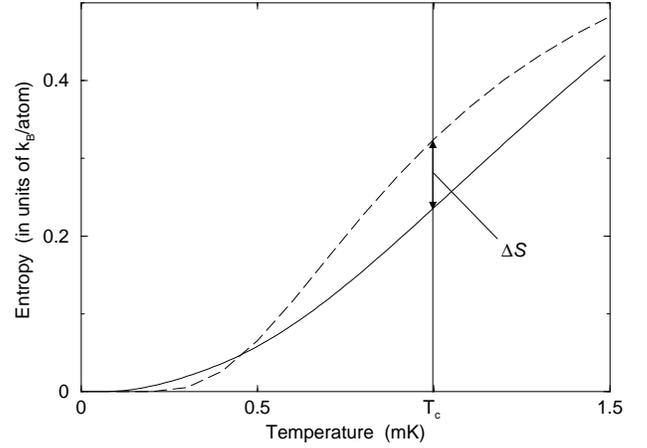}} 
\caption{The calculated entropies of the ordered phase (solid line) and of the PP (dashed-line). At the approximate transition temperature of 1mK we show the jump in entropy which gives a latent heat at the first-order transition of $\sim 100\mu$K per atom.} 
\end{figure}  
 
\begin{figure}[tbp] 
\centerline{\ \epsfysize 7.5cm \epsfbox{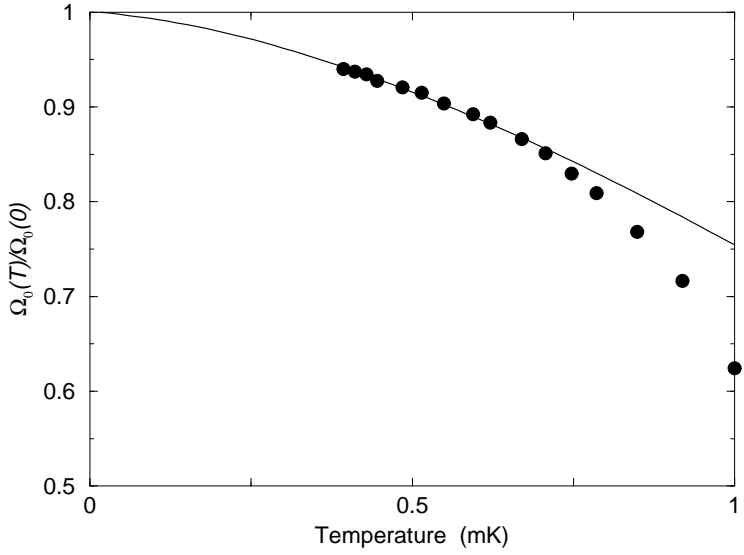}} 
\caption{The temperature dependence of the calculated ordered nuclear magnetic moment (Eq.19) compared with the measured \cite{osheroff} normalized 
antiferromagnetic resonance frequency $\Omega_{0}$.} 
\end{figure}  
 
\begin{figure}[tbp] 
\centerline{\ \epsfysize 7cm \epsfbox{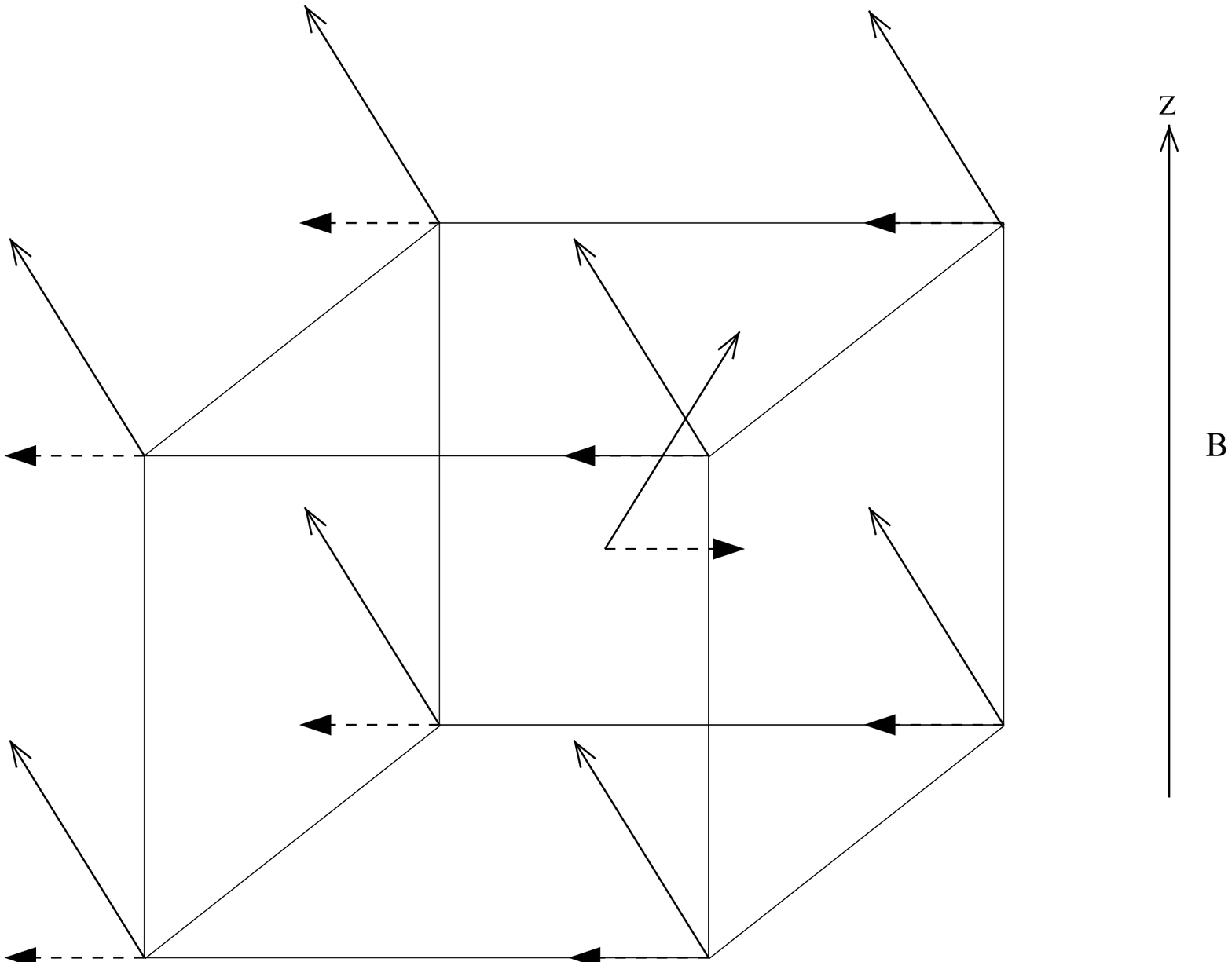}} 
\caption{The High-Field Phase (HFP) arrangement of the nuclear spins. A ferromagnetic arrangement in each sublattice of the $xy$-component (dashed arrows), with zero total $xy$ magnetization. The external field $B$ is in the $z$-direction.} 
\end{figure} 
 
\begin{figure}[tbp] 
\centerline{\ \epsfysize 5cm \epsfbox{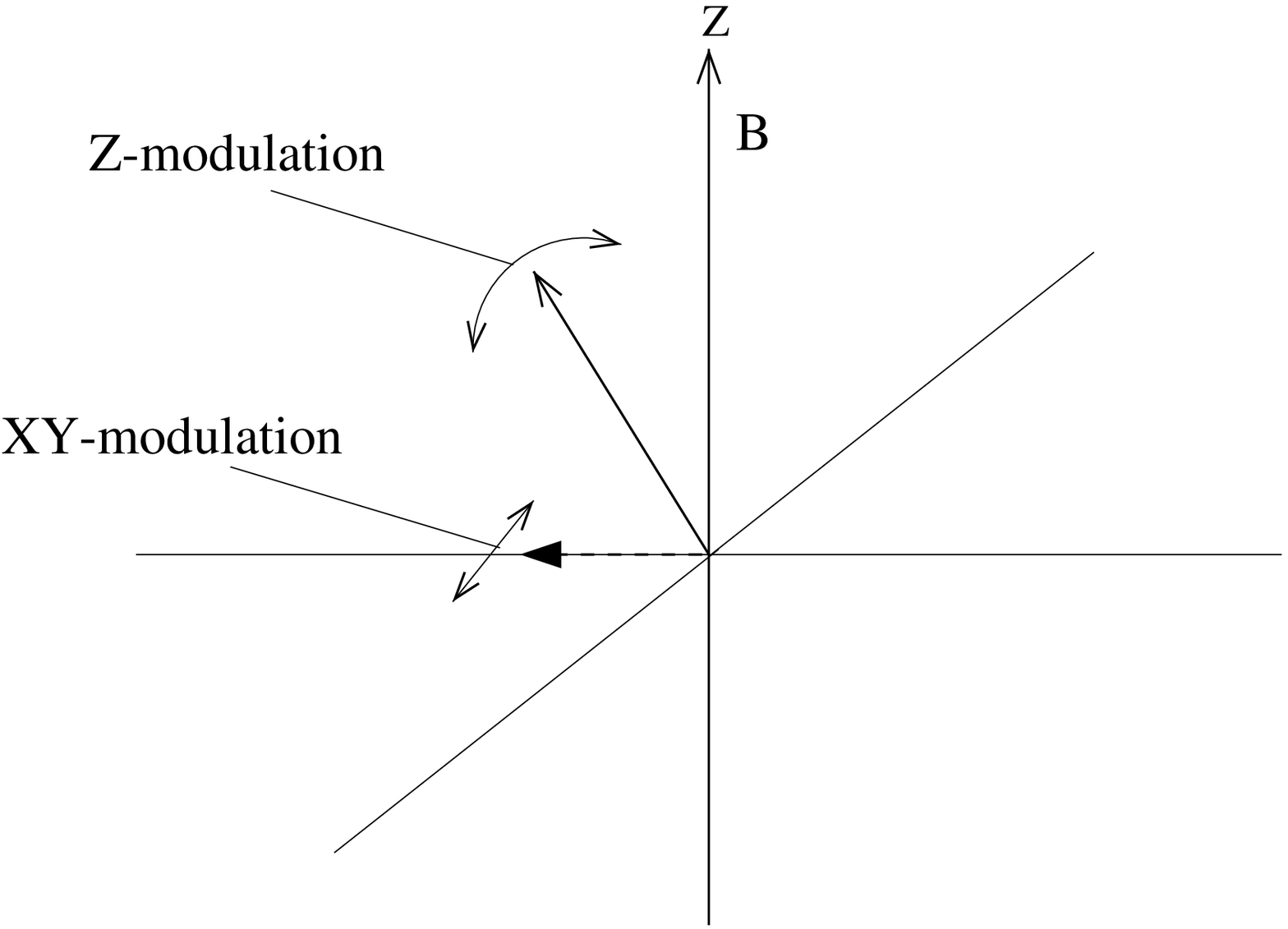}} 
\vskip 0.1cm  
\centerline{\ \epsfysize 7cm \epsfbox{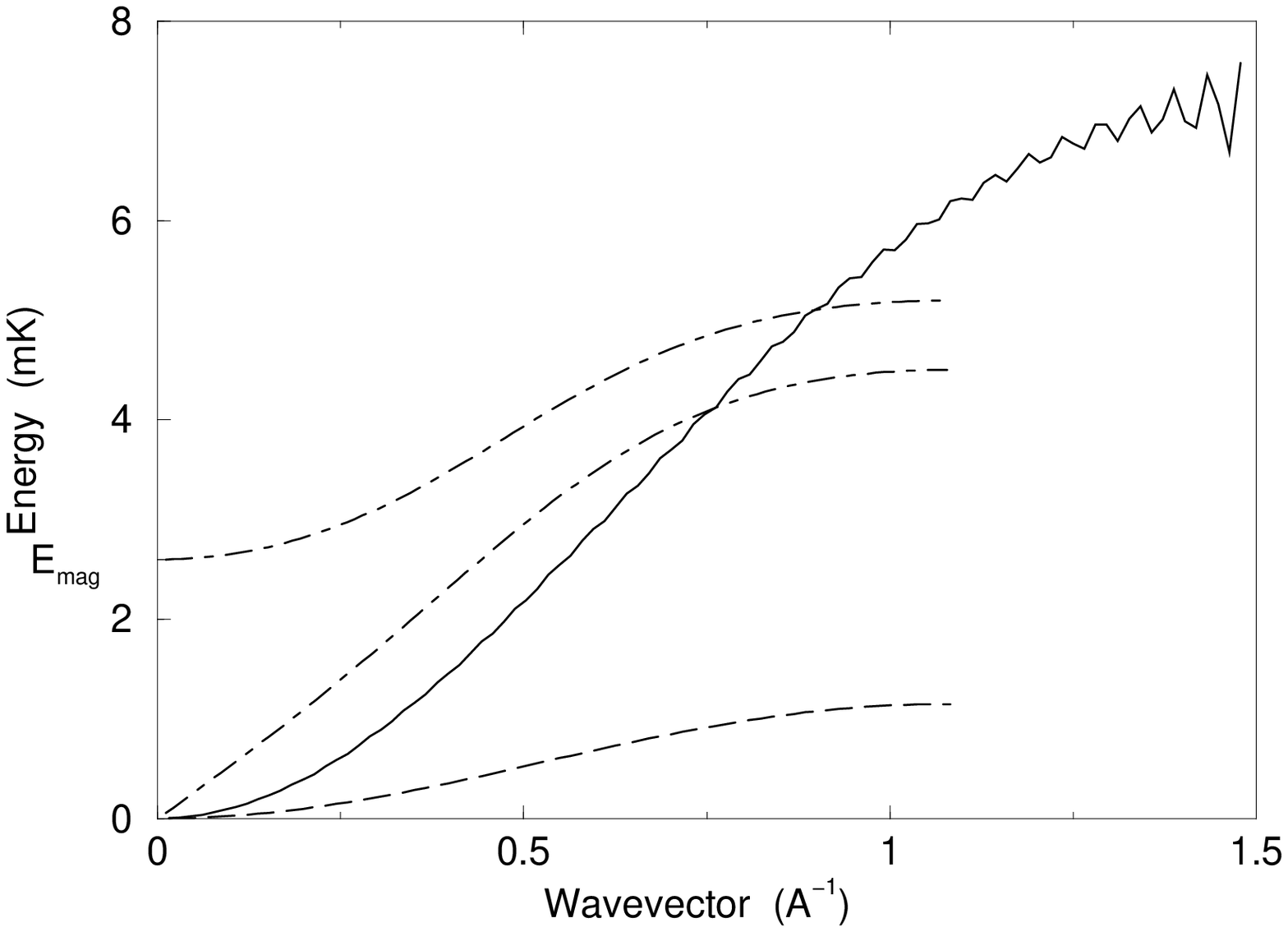}} 
\caption{(a) Schematic description of the small spin perturbations of the $xy$ and $z$ components. (b) The dipolar interaction matrix $X_{hfp}(k)$ between the $xy$ electronic dipoles ${\bf \mu }_{m}$, as a function of the wavevector in the (001)-direction (solid line), (101)-direction (dashed line) and typical spin-waves of the u2d2 phase (dash-dot lines).} 
\end{figure} 
\newpage 
\begin{figure}[tbp] 
\centerline{\ \epsfysize 7cm \epsfbox{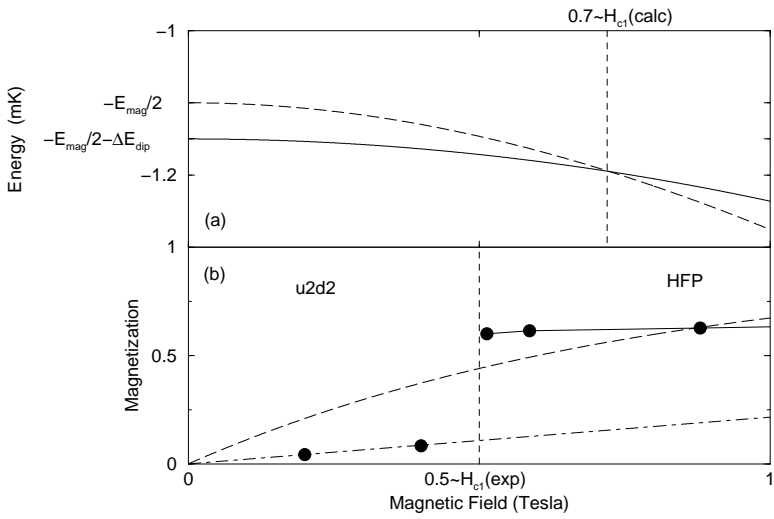}} 
\caption{(a) Comparison between the calculated zero-temperature energies of the u2d2 phase ($E_{u2d2}$ Eq.\ref{freeu2}, solid line), HFP ($E_{hfp}$ Eq.\ref{ehfp}, dashed line) as a function of magnetic field. The calculated transition at $H_{c1}\simeq 0.7$T. (b) Comparison between the calculated zero-temperature magnetization of the u2d2 phase ($N_{u2d2}$, dash-dot line), HFP ($N_{hfp}$, dashed line) and the experimental data \cite{osherof87} (solid circles with guiding solid line). The measured transition is at $H_{c1}(exp)\sim0.5$T.} 
\end{figure} 
 
\begin{figure}[tbp] 
\centerline{\ \epsfysize 7cm \epsfbox{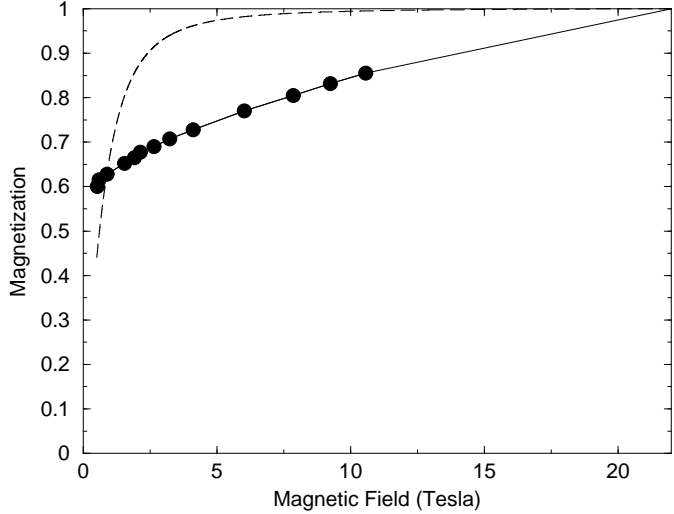}} 
\caption{Comparison between the calculated zero-temperature magnetization of the HFP ($N_{hfp}$, dashed line) and the experimental data \cite{osherof87} (solid circles with solid line).} 
\end{figure} 
 
\begin{figure}[tbp] 
\centerline{\ \epsfysize 7cm \epsfbox{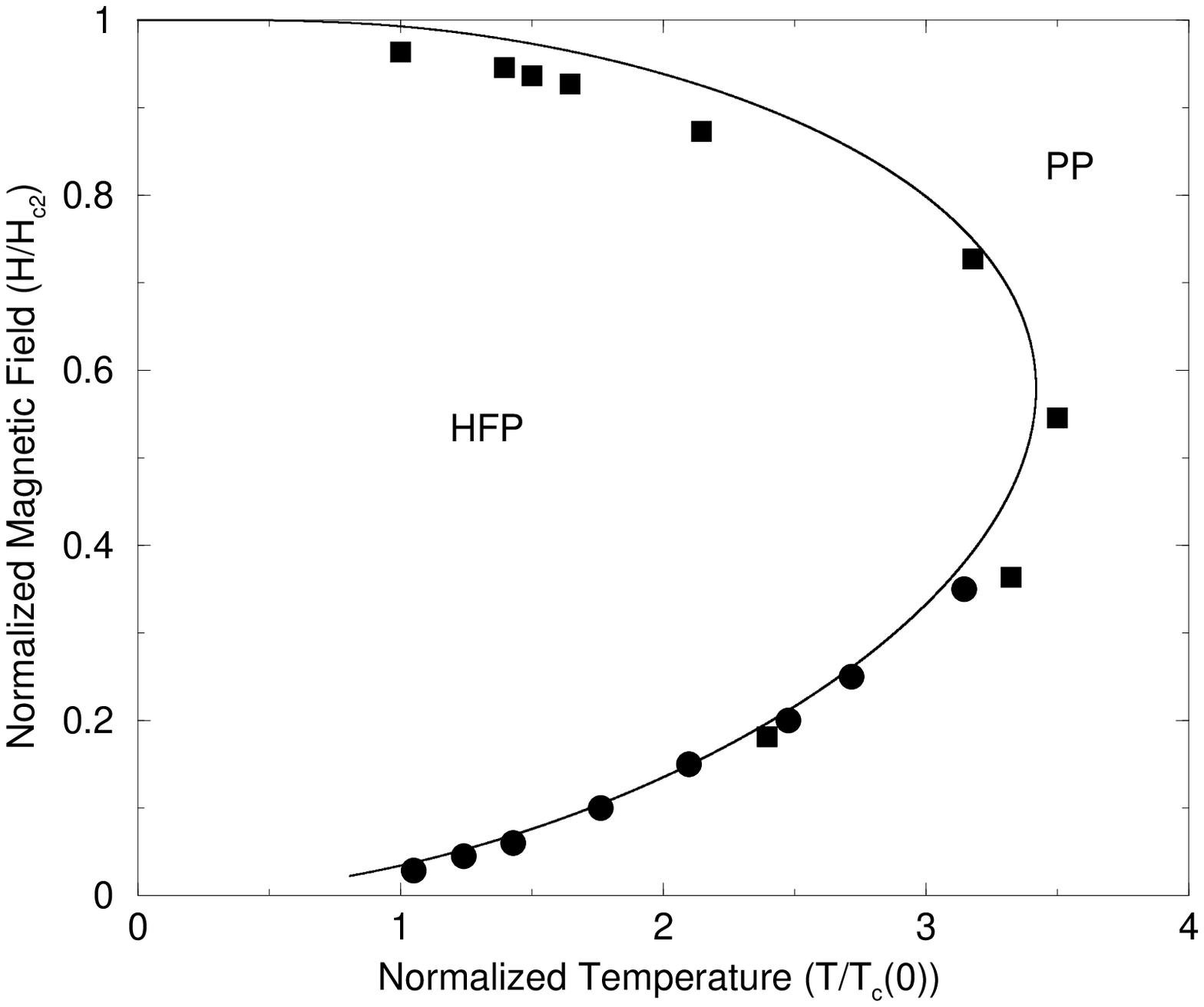}} 
\caption{The geometric-mean energy $E_{mean}\times 1.3$ (solid line) compared with the normalized transition temperature $T/T_{c}(0)$ (solid circles \cite{grey1}, solid squares \cite{fukuyama}) as a function of normalized magnetic field $H/H_{c2}$.} 
\end{figure} 
 

\begin{references} 
 
\bibitem{fisher}  M. Cross and D. Fisher, {\em Rev. Mod. Phys.} {\bf 57} (1985) 881. 
 
\bibitem{cross}  M. Cross, ${\em Jap.\ J. of\ App.\ Phys.}${\it \ }{\bf 26 }%
(1987) 1855. 
 
\bibitem{halperin}  W.P. Halperin, F.B. Rasmussen, C.N. Archie and R.C. 
Richardson, ${\em J.\ Low\ Temp.\ Phys.}${\it \ }{\bf 31 }(1978) 617. 
 
\bibitem{roger}  M. Roger, J.H. hetherington and J.M. Delrieu, {\em Rev. 
Mod. Phys.} {\bf 55} (1983) 1. 
 
\bibitem{grey1}  D.S. Greywall and P.A. Busch, {\em Phys. Rev. B}{\it \ }%
{\bf 36} (1987) 6853. 
 
\bibitem{ceperley}  D.M. Ceperley and G. Jacucci, {\em Phys. Rev. Lett.}{\it %
\ }{\bf 58} (1987) 1648. 
 
\bibitem{cowan}  B. Cowan and M. Fardis, {\em Low Temp. Phys.}{\it %
\ }{\bf 23} (1997) 448. 
 
\bibitem{kumar}  J. DuBois and P. Kumar, {\em J. Low Temp. Phys.} {\bf 98} (1995) 37. 
 
\bibitem{osherof88}  D.D. Osheroff and H. Godfrin, {\em Phys. 
Rev. B} {\bf 38} (1988) 4492. 
 
\bibitem{qfs}  N. Gov and E. Polturak, QFS2000 conference, to appear in {\em J. Low Temp. Phys.}. 
 
\bibitem{niremil}  N. Gov and E. Polturak, {\em Phys. Rev. B }{\bf 60} 
(1999) 1019. 
 
\bibitem{nirbcc}  N. Gov, {\em Phys. Rev. B } (2000) {\bf 62} 
(2000) 910. 
 
\bibitem{suzuki}  Y. Masuda and H. Suzuki, {\em J. Low Temp. Phys.}{\it \ }%
{\bf 75} (1989) 159. 
 
\bibitem{glyde}  H.R. Glyde, 'Excitations in Liquid and Solid Helium', 
Oxford Series on Neutron Scattering in Condensed Matter, 1994. 
 
\bibitem{nosanow}  L.H. Nosanow {\em Phys. Rev.} {\bf 146} (1966) 120. 
 
\bibitem{vibronic}  I.B. Bersuker and V.Z. Polinger, `Vibronic Interactions in Molecules and Crystals', 1983, Springer Series in Chemical Physics 49. 
 
\bibitem{goodkind}  C.A. Burns and J.M. Goodkind {\em J. Low Temp. Phys.} {\bf 95} (1994) 695. 
 
\bibitem{greywall}  D.S. Greywall, {\em Phys. Rev. A}{\it \ }{\bf 3} (1971) 
2106. 
 
\bibitem{kohler}  T.R. Kohler, {\em Phys. Rev. Lett.}{\it \ }{\bf 18} (1967) 
654. 

\bibitem{heald}  S.M. Heald, D.R. Baer and R.O. Simmons, {\em Phys. Rev. B}%
{\it \ }{\bf 30} (1984) 2531. 
 
\bibitem{grey}  D.S. Greywall, {\em Phys. Rev. B}{\it \ }{\bf 15} (1977) 
2604. 
 
\bibitem{izumi}  I. Iwasa and H. Suzuki, ${\em J.\ Low\ Temp.\ Phys.}${\it \ 
}{\bf 62 }(1986) 1. 
 
\bibitem{simmons}  C. Seyfert, R.O. Simmons, H. Sinn, D.A. Arms and E. Burkel {\em J. Phys.: Condens. Matter} {\bf 11} (1999) 3501. 
 
\bibitem{cohen}  Cohen Tanoudji, 'Quantum Mechanics', Vol. II, 
Wiley-Interscience (1977). 
 
\bibitem{xxzmodel} R.F. Bishop {\em et. al.}, cond-mat/0011008. 

\bibitem{hyperx}  W. Vassen and W. Hogervorst, {\em Phys. Rev. A} {\bf 39} (1989) 4615. 
 
\bibitem{ashcroft}  N.W. Ashcroft and N.D. Mermin, 'Solid State Physics', 
Saunders College Publishing, 1976. 

\bibitem{fuku91}  H. Fukuyama {\em et. al.}, {\em Phys. Rev. Lett.} {\bf 67} (1991) 1274. 
 
\bibitem{osheroff}  D.D. Osheroff, M.C. Cross and D.S. Fisher, {\em Phys. 
Rev. Lett.}{\it \ }{\bf 44} (1980) 792. 
 
\bibitem{bossy}  A. Benoit, J. Bossy, J. Flouquet and J. Schweizer, ${\em %
J.\ Phys.\ Lett.}${\it \ }{\bf 46 }(1985) L923. 
 
\bibitem{okamoto}  T. Okamotoa {\em et. al.}, {\em Phys. Rev. Lett.} {\bf 71} (1994) 868. 
 
\bibitem{hcpdipole} The measured ferromagnetic short-range correlations \cite{okamoto,lang} in the hcp phase can result 
from nearest-neighbor (i.e. local) dynamical correlations. 
Local correlations of the zero-point motion will result in electronic 
polarization and hyperfine interactions similar in size to those calculated 
for the bcc. 
These may cause the observed local (i.e. randomly oriented) ferromagnetic correlations.
 
\bibitem{anderson}  P.W. Anderson, 'Concepts In Solids', 1963, p.132-148. 
 
\bibitem{huang}  K. Huang, 'Statistical Mechanics',1987, John Wiley and 
Sons,Inc.. 
 
\bibitem{osherofyu}  D.D. Osheroff and C. Yu, {\em Phys. Lett.} {\bf 77A} (1980) 458. 
 
\bibitem{sun}  Z. Sun and J.H. Hetherington, {\em J. Low Temp. Phys.} {\bf 91} (1993) 299. 
 
\bibitem{fuku87}  H. Fukuyama {\em et. al.}, {\em Phys. Rev. B} {\bf 36} (1987) 8921. 
 
\bibitem{kittel}  Kittel, 'Introduction to Solid State Physics'. 

\bibitem{osherof87}  D.D. Osheroff, H. Godfrin and R. Ruel, {\em Phys.  Rev. Lett.} {\bf 58} (1987) 2458.  
\bibitem{osherof90}  Y.P. Feng, P. Schiffer, J. Mihalisin and D.D. Osheroff, {\em Phys. 
Rev. Lett.} {\bf 65} (1990) 1450. 
 
\bibitem{sasaki}  Y. Sasaki, T. Matsushita, T. Mizusaki and A. Hirai, {\em Phys. 
Rev. B} {\bf 44} (1991) 7362. 
 
\bibitem{hfp}  D.D. Osheroff, {\em Physica B} {\bf 109,110} (1982) 1461. 
 
\bibitem{domb}  C. Domb and M.S. Green, 'Phase Transitions and Critical Phenomena', Vol.3, p. 570, 
 London, New York, Academic Press (1974). 
 
\bibitem{fukuyama}  H. Fukuyama {\em et. al.}, {\em Physica B} {\bf 169} (1991) 197. 
 
\bibitem{patria}  R.K. Patria, 'Statistical Mechanics',1984, Pergamon Press Ltd., Oxford, England. 
 
\bibitem{okamoto}  T. Okamotoa {\em et. al.}, {\em Phys. Rev. Lett.} {\bf 71} (1994) 868.

\bibitem{lang}  T. Lang {\em et. al.}, {\em Phys. Rev. Lett.} {\bf 77} (1996) 322.

\end{references}
\end{document}